\documentclass[aps,prd,twocolumn,groupedaddress,showpacs,nofootinbib]{revtex4}
\usepackage{amsmath,amssymb}
\usepackage{graphicx}
\begin{document}
\title{Hidden past of dark energy cosmological models}
\author{L. Fern\'andez-Jambrina}
\email[]{leonardo.fernandez@upm.es}
\homepage[]{http://debin.etsin.upm.es/ilfj.htm}
\affiliation{Matem\'atica Aplicada, E.T.S.I. Navales, Universidad
Polit\'ecnica de Madrid,\\
Arco de la Victoria s/n, \\ E-28040 Madrid, Spain}
      
\begin{abstract}In this paper we analyse the possibility of having
homogeneous isotropic cosmological models with observers reaching
$t=\infty$ in finite proper time.  It is shown that just
observationally-suggested dark energy models with $w\in(-5/3,-1)$ show
this feature and that they are endowed with an exotic curvature
singularity.  Furthermore, it is shown that non-accelerated observers
in these models may experience a duration of the universe as short as
desired by increasing their linear momentum.  A subdivision of phantom
models in two families according to this behavior is suggested.
\end{abstract} 

\maketitle

\section{Introduction}

During the last years there has been mounting experimental evidence
from different sources (supernovae type Ia \cite{supernova}, redshift
of distant objects \cite{redshift} and temperature fluctuations of
background radiation \cite{cmbr}) supporting an accelerated expansion
of our Universe at present time (cfr.  for instance \cite{de} for a
review).

Trying to explain this fact, several proposals have been made, such as
dark energy contents for the universe or modifications of the theory
of gravity, which have produced a menagerie of new types of singular
events in the respective cosmological models, traditionally restricted
to Big Bang and Big Crunch singularities.  For instance, we may find
in phantom energy models Big Rip singularities \cite{caldwell}. One 
of this models has been shown to be stable against quantum 
corrections \cite{kahya}. An
attempt to explain the accelerated expansion without violating all energy
conditions \cite{sudden} produces sudden singularities. Most 
recently, inaccessible singularities in toral cosmologies have been 
added to the list \cite{mcinnes}.

There have been several attempts to organize these families of
singular events in thorough classifications.  In \cite{visser} all
types of singular events in Friedmann-Lema\^{\i}tre-Robertson-Walker
(FLRW) cosmological models are classified according to the
coefficients and exponents of a power expansion in time of the scale
factor of the universe around the event.  In \cite{classodi}
singularities are classified using the finiteness of the scale factor,
the density and the pressure of the universe.  In \cite{puiseux} the
behavior of causal geodesics close to singular events and the strength
of the singularities are analysed.

This line of research has proven successful showing unexpected features
of FLRW cosmological models near the singularities.  For instance, it
has allowed to show that sudden singularities are weak \cite{flrw},
since tidal forces do not disrupt finite objects falling into them
\cite{ellis,tipler, krolak}.

Another intriguing feature concerning Big Rip singularities is that
photons do not experience such fate for effective equations of state,
$p=w\rho$, with $w\in(-5/3,-1)$ (that is, those comprised between the
superphantom \cite{dabrowski} and the phantom divide), since they
require an infinite lapse of time to reach that event \cite{puiseux}.
Since this range of the parameter $w$ comprises the observationally
accepted values \cite{obs}, which are slightly below -1, this fact is
more than a mere curiosity.

Following the idea of classifying the singular events arising in FLRW
cosmological models, it is worth mentioning that all classifications
are incomplete in a sense: they unveil what happens at a finite
coordinate time $t$, but they are elusive when asked about infinite
$t$.  This may seem a pointless consideration, since in most cases an
infinite coordinate time lapse corresponds to an infinite time lapse
experienced by the observer, but the mentioned example about photons in
phantom cosmologies, where a finite coordinate time lapse requires an
infinite proper time shows us that the issue is far from being
trivial.

To this aim in Sec. \ref{eqs} the equations governing causal 
geodesics in FLRW cosmological models are reviewed. In Sec. 
\ref{infgeod} the conditions for a causal geodesic to reach $t=\pm 
\infty$ in finite proper time are derived. It will be shown that just 
phantom models fulfill this property. In Sec. \ref{curva} it will be 
discussed if this abrupt end of causal geodesics is an actual 
singularity or not. Analysis of the Ricci curvature as measured by 
the observers will settle the issue, in spite of the zero value of 
curvature scalar polynomials there. In fact these are strong 
curvature singularities. Finally, the consequences of these facts 
will be discussed in Sec. \ref{discuss}.

\section{Geodesics in FLRW cosmological models\label{eqs}}

The metric for FLRW cosmological models may be written,
\begin{eqnarray}
&&ds^2=-dt^2+a^2(t)\left\{f^2(r)dr^2+ r^2\left(d\theta^2+\sin^2\theta
d\phi^2\right)\right\}\nonumber\\
&&f(r)=\frac{1}{\sqrt{1-kr^2}},\quad k=0,\pm1,\label{metric}\end{eqnarray}
in terms of spherical coordinates $r,\theta,\phi$ with their usual 
ranges and a coordinate time, with a range depending on the type of
cosmological model. 

Three families of models are comprised in this 
expression, open models with $k=-1$, flat models with $k=0$ and 
closed models with $k=1$. Observations favor flat models, but we keep 
for our purposes the general formula.

Free-falling observers in a spacetime are modeled by timelike
geodesics parametrized by proper time $\tau$, since these curves have
the property of vanishing acceleration.  The use of proper time allows
us to write the velocity $u$ of the parametrization of the geodesic
$(\dot t, \dot r, \dot\theta, \dot\phi)$ as a unitary vector,
\begin{equation}\label{delta}
\delta=-g_{ij}\dot x^{i}\dot x^{j}, \quad x^i,
x^j=t,r,\theta,\phi,
\end{equation}
where the dot means derivation with respect to $\tau$.

There are three types of geodesics: timelike ($\delta=1$), spacelike
($\delta =-1$) and lightlike ($\delta=0$).  We consider just causal
geodesics, $\delta=0,1$, since they are the only ones that may carry
signals or observers.

A quick way to write down simple geodesic equations for these
spacetimes is taking into account that the universe is homogeneous 
and isotropic and therefore geodesics are straight lines in the 
spacetime and we may take
$\dot \theta=0=\dot \phi$ without loss of generality.

It is also easy to check that the vector
$\partial_{R}=\partial_{r}/f(r)$,
\[R=\left\{
\begin{array}{ll}
    \mathrm{arcsinh}\, r & k=-1  \\
    r & k=0  \\
    \arcsin r & k=1
\end{array}
\right.,\]generates an isometry along these
straight lines and therefore there is a conserved quantity $P$ of 
geodesic motion attached to it, the specific linear momentum of the 
observer,
\begin{equation} \pm P=u\cdot\partial_{R}=a^2(t)f(r)\dot
r,\end{equation}where the $\cdot$ denotes the inner product defined 
by the metric (\ref{metric}).  The double sign is introduced in order 
to keep $P$ positive.

We just need another equation for $\dot t$ to complete the set and 
we may obtain it without resorting to Christoffel symbols for the 
metric by using the unitarity condition (\ref{delta}),
\[\delta =\dot t^2-a^2(t)f^2(r) \dot r^2.\]

Restricting to future-pointing geodesics, $\dot t>0$ (past-pointing 
geodesics are treated in a similar fashion), the whole set of 
geodesic equations is reduced to
\begin{subequations}\begin{eqnarray}\label{geods}
    \dot t&=&\sqrt{\delta +\frac{P^2}{a^2(t)}},\label{geods1}\\\dot
    r&=&\pm\frac 
    {P}{a^2(t)f(r)}.\label{geods2}\end{eqnarray}\end{subequations}

Hence we see that there are basically three types of causal geodesics:
radial lightlike ($\delta=0$, $P\neq 0$) and timelike geodesics 
($\delta=1$, $P\neq 0$) and the comoving congruence of
fluid worldlines ($\delta=1$, $P= 0$), which provide little
information about the geometry of spacetime, since for them $t=\tau$
regardless of the possible singularities in the universe.

\section{Singularities at infinity\label{infgeod}}

Since singularities along causal geodesics at a finite $t_{0}$ were 
considered in detail in \cite{puiseux}, we focus now on infinite 
values of coordinate time $t$.

Singularities may appear also at $t=\pm \infty$ if there are observers
that reach these events in finite proper time.  Unfortunately, it is
not always possible to perform power expansions of the scale factor
centered in $t=\pm \infty$, as it is done in \cite{visser, 
puiseux} for finite $t$, since there are physically reasonable
spacetimes with oscillatory scale factors, for instance, anti-de
Sitter universes, for which the limit of $a(t)$ is not defined when
$t$ tends to infinity.

However, the question of when $t=\pm\infty$ is reached by geodesic
observers in finite proper time can be easily solved.

For lightlike radial geodesics we have
\[ \dot t=\frac{P}{a(t)},\]
\[\int_{t_{0}}^t a(t')\,dt'=P(\tau-\tau_{0}),\]
and therefore lightlike geodesics reach $t=\infty$ in finite proper time
if and only if the integral
\begin{equation}\label{conda}\int_{t}^\infty a(t')\,dt'\end{equation}
is finite for sufficiently large $t$. That is, if $a(t)$ is an integrable function at infinity.

Comoving fluid wordlines with $P=0$ need not be considered, since they
reach $t=\infty$ in infinite proper time.  

Finally, we have timelike radial geodesics.  In this
case, proper time may be written again in terms of an integral of
$a(t)$ using (\ref{geods1}), \begin{equation}\label{duration}\int_{t_{0}}^t
\frac{dt'}{\sqrt{1+P^2/a^2(t')}}=\tau-\tau_{0},\end{equation}and therefore these
geodesics reach $t=\infty$ in finite proper time if and only if the
improper integral \[\int_{t}^\infty \frac{dt'}{\sqrt{1+P^2/a^2(t')}}\]is
convergent for sufficiently large $t$.

Obviously this can only happen if $a(t)$ tends to zero at infinity,
but it is not a sufficient condition.  Since we may bound
\[\int_{t}^\infty \frac{dt'}{\sqrt{1+P^2/a^2(t')}}<\frac{1}{P}\int_{t}^\infty
a(t')dt',\]the integral for timelike geodesics is convergent if the
one for lightlike geodesics is.  

Furthermore, since for large $t$ and
$a(t)$ tending to zero \[\frac{1}{\sqrt{1+P^2/a^2(t')}}=
\frac{a(t')}{P}-\frac{1}{2}\frac{a^3(t')}{P^3}+\cdots,\]is a
telescopic series, the integral for timelike geodesics converges if
and only if the one for lightlike geodesics does.  Hence, all radial
geodesics have the same regularity pattern.

The analysis for $t=-\infty$ is entirely similar and so we have
focused on the $t=\infty$ case.

Since in most models the scale factor $a(t)$ behaves asymptotically as
a power of coordinate time, we start considering scale factors which
behave close to infinity as \[a(t)\simeq c|t|^{\eta}, \quad c>0, \quad
w=\frac{2}{3\eta}-1.\]

The equation for lightlike geodesics (\ref{geods1}) may be integrated close to
infinity,
\begin{eqnarray*}t&\simeq&\left\{\frac{(1+\eta) P}{c}\right\}^{1/(1+\eta)}
(\tau-\tau_{0})^{1/(1+\eta)},\quad \eta\neq-1,\\
t&\simeq&e^{P(\tau-\tau_{0})/c},\quad \eta=-1,\end{eqnarray*}
and provides valuable information, since $t$ diverges when $\tau$
tends to infinity for $\eta\ge-1$, whereas $t$ diverges at finite proper
time $\tau_{0}$ if $\eta<-1$.

The latter cases are quite interesting, since at $\tau_{0}$ the
geodesic reaches $t=\infty$ in finite proper time.  Therefore,
lightlike geodesics range from $\tau=-\infty$ ($t=0$) to
$\tau=\tau_{0}$ ($t=\infty$) and are incomplete towards the future. 

This is not the interesting case, since it involves models starting at
a Big Rip at $t=0$.  But If we consider $t=-\infty$, lightlike geodesics range
from $\tau=\tau_{0}$ ($t=-\infty$) to $\tau=\infty$ ($t=0$) and are
incomplete towards the past.  This is the usual range in the suggested
phantom models.

As it has been said, the same behavior appears for timelike radial 
geodesics, with the difference that these actually end up at the Big Rip 
at $t=0$ in a finite proper time \cite{puiseux}.

%
%
%
%
%
%
%
 
Not only causal geodesics, but also spatial geodesics show this 
feature.

For non-tilted spatial geodesics in a hypersurface $t=t_{0}$,
\[\dot t=0\Rightarrow t=t_{0},\quad P=a(t_{0}),\]
\[\dot r=\pm\frac{1}{a(t_{0})f(r)}\Rightarrow R=\pm 
\frac{s-s_{0}}{a(t_{0})},\]
proper distance $s$ is essentially the radial coordinate $R$, 
corrected by the expansion factor, as 
expected.

But for tilted spatial geodesics,
\[\dot t=\sqrt{\frac{P^2}{a^2(t)}-1}\Rightarrow 
s-s_{0}=\int^\infty_t\frac{dt'}{\sqrt{P^2/a^2(t')-1}},\]
and for $a(t)\simeq c|t|^\eta$ for large $t$, this integral converges to 
a finite value if and only if $\eta<-1$. Hence the length of these 
tilted spatial geodesics is also finite, even though the radial 
coordinate $r$ diverges.

\section{Curvature singularities\label{curva}}

However, at $t=\pm\infty$ all curvature scalar polynomials vanish,
since they decrease as $t^{-2}$ and this suggests a sort of
Minkowskian limit.  Therefore there is no scalar polynomial curvature
singularity there. A pathological feature named imprisoned 
incompleteness, which appears in spacetimes like Taub-NUT, where 
geodesics are incomplete without singular scalars of curvature, is 
not feasible, since the spacetime has a cosmic time (a function with 
timelike gradient everywhere, the coordinate time $t$) and is 
therefore causally stable \cite{HE,seno}. We might suspect that geodesic
incompleteness could simply point out that the spacetime is not fully
covered with the coordinate patch (\ref{metric}) and that therefore it
could be extendible beyond $t=\pm\infty$.

This is the case, for instance, of Milne universe, corresponding to
$k=-1$, $a(t)=t$ in (\ref{metric}).  A suitable coordinate
transformation \[T=t\sqrt{1+r^2},\quad R=rt,\] shows that this
model is just the portion of Minkowski spacetime inside the null cone
$T=R$ and therefore the apparent singularity at $t=0$ is due just to
the choice of coordinates.  

A similar feature exhibits de Sitter spacetime in the parametrization
that is usually used for inflation, $k=0$,
$a(t)=e^{\sqrt{\Lambda/3}\,t}$, which fulfils condition (\ref{conda})
and therefore its radial geodesics reach $t=-\infty$  in finite 
proper time. However, again in this case it is possible to extend the 
spacetime to a larger one, $k=1$, 
$a(T)=\sqrt{3/\Lambda}\cosh\left(\sqrt{\Lambda/3}\,T\right)$ with 
another change of coordinates \cite{HE} and hence the singularity at 
$t=-\infty$ is only apparent.

Or for Schwarzschild spacetime, which in
Schwarzschild coordinates appears to be singular at the horizon at
$r=2M$, whereas this coordinate singularity disappears on extending it
with Eddington-Finkelstein \cite{eddington} or Kruskal
\cite{kruskal} coordinates.

However, the null value of the scale factor in that limit suggests a 
point as a limit in this case. 

If we compute the Ricci tensor component along the velocity of the
geodesic, an exotic behavior appears.

For a radial lightlike geodesic,
\[u^t=\dot t=\frac{P}{a},\qquad u^r=\dot r=\pm\frac{P}{fa^2},\]
\begin{equation}\label{ricci}
R_{ij}u^iu^j=2P^2\left(\frac{a'^2+k}{a^4}-\frac{a''}{a^3}\right)\simeq
\frac{2P^2\eta}{c^2t^{2(\eta+1)}}
+\frac{2kP^2}{c^4t^{4\eta}},\end{equation}
we take a look at the first term of the zero component of the Ricci 
curvature, since it is present regardless
of the value of $k$,
\begin{equation}2P^2\left(\frac{a'^2}{a^4}-\frac{a''}{a^3}\right)\simeq
\frac{2P^2\eta}{c^2t^{2(\eta+1)}}\simeq
\frac{2\eta}{(\eta+1)^2}\frac{1}{(\tau-\tau_{0})
^{2}},\label{sign}\end{equation}
and find out that the Ricci curvature diverges when $t$ approaches $\pm 
\infty$ ($\tau$ tends to $\tau_{0}$) for $\eta<-1$.

The result invoked for the singularity-free de Sitter spacetime, 
$k=0$, $a(t)=e^{\sqrt{\Lambda/3}\,t}$ is
also recovered since in this case the expression for the Ricci
curvature along the geodesics is zero,
\[a'^2+k=aa''.\]

The remaining solutions for this equation are other parametrizations of the de
Sitter spacetime, $k=1$,
$a(t)=\sqrt{3/\Lambda}\cosh\left(\sqrt{\Lambda/3}\,t\right)$; $k=-1$,
$a(t)=\sqrt{3/\Lambda}\sinh\left(\sqrt{\Lambda/3}\,t\right)$ and the
anti-de Sitter spacetime, $k=-1$,
$a(t)=\sqrt{3/\Lambda}\cos\left(\sqrt{\Lambda/3}\,t\right)$, but for 
a choice of time coordinate origin. None of them are decreasing at 
$t=\pm\infty$, so they do not affect our results.

A similar analysis may be performed for radial timelike geodesics,
\[u^t=\dot t=\sqrt{1+\frac{P^2}{a^2}},\qquad u^r=\dot
r=\pm\frac{P}{fa^2},\] 
\begin{equation}\label{ricci1}R_{ij}u^i
u^j=-\frac{3a''}{a}+2P^2\left(\frac{a'^2+k}{a^4}-\frac{a''}{a^3}
\right),\end{equation} since for scale factors with $\eta<-1$ and large values of $|t|$
the dominant term is the $P^2$-term, which is the same as for
lightlike geodesics.  

Hence the Ricci curvature diverges along 
both families of radial geodesics on approaching $t=\pm \infty$.

Hence we are to conclude that universes with scale factor $a(t)\simeq
c|t|^{\eta}$, $\eta<-1$, for large values of $|t|$ have a p.p. curvature
singularity (curvature singularity along a parallelly transported
basis) \cite{HE} at $t=\pm \infty$, though the scalar polynomials of
curvature are zero there. 

That is, there is an \emph{actual} curvature 
singularity at $t=\pm \infty$ for these models, which correspond to 
$w\in(-5/3,-1)$, which is reached by the observers in finite proper 
time. Considering just expanding models of this type, all radial observers would trace their geodesic paths from the 
initial singularity at $t=-\infty$ to the Big Rip at $t=0$ in a 
finite lapse of proper time.

This result does not contradict the Penrose diagrams for these models
shown in \cite{chiba}, since conformal diagrams provide no 
information about distances, just about angles, but introduces a 
difference between models with $w\in (-5/3,-1)$ and those with $w\le 
-5/3$ as it is shown in Fig. 1.

    \begin{figure}[h]\begin{center}
    \includegraphics[height=0.2\textheight]{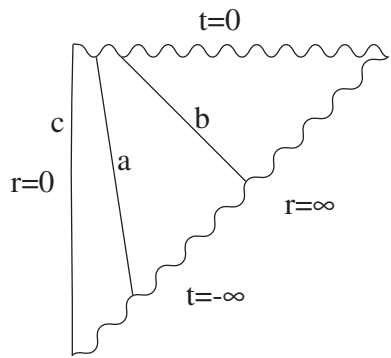}
    \caption{Conformal diagram for a model with $w\in (-5/3,-1)$: 
    Timelike radial geodesics like $a$ have finite length, whereas 
    lightlike geodesics like $b$ are infinite towards the future and 
    timelike geodesics like $c$ are infinite towards the past.}
    \label{penrose}\end{center}
    \end{figure}

Furthermore, we may check the strength of these curvature
singularities, which might be relevant, since other types of
singularities, such as sudden singularities \cite{sudden} 
(singularities II and IV in \cite{classodi}) were shown not to be strong enough
to disrupt finite objects \cite{flrw} and even have been suggested to
be consistent with observations \cite{hendry}.

Definitions of singularities related to curvature and geodesics refer 
to ideal point observers. When finite objects are considered, tidal 
forces are relevant and it is interesting to check if they may destroy the 
object. In this case the singularity is considered to be strong 
\cite{ellis}. This qualitative concept has been stated rigorously by 
several authors \cite{tipler,krolak,krorud,rudnicki}.

For instance, in Tipler's definition \cite{tipler} a curvature
singularity is strong if the volume spanned by three Jacobi fields
refered to a orthormal basis parallely-transported along a causal
geodesic tends to zero at the singularity.  Kr\'olak's definition
\cite{krolak} just requires that the derivative of this volume be
negative.

There are necessary and sufficient conditions for the appearance of 
strong singularities \cite{clarke}, that become quite simple to 
implement in the case of FLRW spacetimes, since the Weyl tensor 
vanishes \cite{puiseux}.

With Tipler's definition a lightlike geodesic of velocity $u$ comes up a
strong singularity at $\tau_{0}$ if and only if the integral
\begin{equation}\label{suftipler}
 \int_{0}^{\tau}d\tau'\int_{0}^{\tau'}d\tau''R_{ij}u^{i}u^j
\end{equation}
diverges as $\tau$ tends to $\tau_{0}$.

And with Kr\'olak's definition a lightlike geodesic of velocity $u$ comes up a strong
singularity at $\tau_{0}$ if and only if
\begin{equation}\label{sufkrolak}
 \int_{0}^{\tau}d\tau'R_{ij}u^{i}u^j
 \end{equation}
diverges as $\tau$ tends to $\tau_{0}$.

Since the Ricci curvature component (\ref{ricci}) diverges as
$1/(\tau-\tau_{0})^2$, the integral of this term provides a
logarithmic divergence at $\tau_{0}$ for $\eta<-1$ with Tipler's 
definition and an inverse power divergence with Kr\'olak's 
definition. Therefore, lightlike geodesics meet a strong singularity at
$t=\pm\infty$ if and only if $\eta<-1$.  The contribution of the curvature
term diverges even faster when present.

The previous conditions on integrals of Ricci components become 
sufficient conditions on dealing with timelike geodesics. Since the 
behavior of Ricci curvature has been shown to be similar for both 
families of radial geodesics for large $|t|$ and $\eta<-1$, we learn 
that also radial timelike geodesics meet a strong curvature 
singularity at $t=\pm \infty$.

We may easily extend this result to non-power law growth/decrease of the
expansion factor:

\begin{itemize}
    \item  For $a(t)$ growing or decreasing as $1/|t|$ or slower,
    radial geodesics reach $t=\pm\infty$ in infinite proper time.

    \item  For $a(t)$ decreasing faster than $1/|t|$, radial
    geodesics reach $t=\pm\infty$ in finite proper time and therefore
    there is an actual strong curvature singularity there, except for 
    de Sitter spacetime.
\end{itemize}

These two cases include all situations for which the scale factor has 
a well-defined limit $t\to\pm\infty$. Oscillatory scale factors may 
be treated directly with condition (\ref{conda}).

\section{Discussion \label{discuss}}

So far we have shown that FLRW cosmological models for which $a(t)$ 
decreases faster than $1/|t|$ for large values of $|t|$ show a strong 
curvature singularity for $t\to\pm\infty$, except for de Sitter 
spacetime. This is the case of phantom models 
with $w\in(-5/3,-1)$, family that includes models compatible with 
observations, since $w$ is estimated to be slightly below minus 
one \cite{obs}.

Since the implications of these results are related to the past of 
the models instead of their future, it might seem a pointless 
discussion, for phantom models are intended to describe the future of 
the universe from now on. In the past other fields such as dust, 
radiation and the cosmological constant would be dominant and would 
prevent the appearance of the exotic curvature singularities 
described here. 

However, even though phantom models are not relevant to study the 
past of the universe, there are still consequences that are 
applicable to our present universe.

We may consider, for instance, the total duration of a  universe 
filled with a phantom field as experienced by a free-falling observer 
(\ref{duration}),
\begin{eqnarray}T&=&\int_{-\infty}^0\frac{dt}{\sqrt{1+P^2/a^2(t)}}=
\int_{-\infty}^0\frac{dt}{\sqrt{1+P^2/c^2t^{2\eta}}}\nonumber\\&=&
\left(\frac{P}{c}\right)^{1/\eta}\int^{\infty}_0\frac{x^\eta dx}{\sqrt{1+x^{2\eta}}}
\nonumber\\&=&-\left(\frac{P}{c}\right)^{1/\eta}\frac{B\left(-\frac{1}{2\eta},\frac{1}{2}+\frac{1}{2\eta}\right)}{2\eta},
\quad \eta<-1,
\label{P}\end{eqnarray}
by the change of variable $x=-(c/P)^{1/\eta}t$, using the hypergeometric 
function Beta.

We already know that this expression for the time span is finite for 
$\eta<-1$, but it can be made as small as desired by taking arbitrary 
large values of the linear momentum of the observer $P$. There is no 
lower bound, nor upper bound, which we know it is infinite for 
non-radial observers.

Though the calculation has been made for the duration of the universe
from the initial singularity to the Big Rip, it is clear that this
result is also valid for the time span from the coincidence moment 
when phantom fields become the dominant component of the content of 
the universe to the Big Rip. That is, non-accelerated observers may 
shorten the time span to the end of the universe at will by 
increasing their linear momentum $P$. 

This feature is exclusive of dark energy models with 
$w\in(-5/3,-1)$, since a negative exponent $\eta$ is required in 
(\ref{P}) for the decreasing behavior of $T$. For models with $w\le
-5/3$, $\eta\in[-1,0)$ the integral (\ref{P}) is divergent, since 
$t=-\infty$ is actually at infinity, but the decreasing behavior is 
also exhibited for finite intervals of time up to the Big Rip 
singularity, though the interest on these models is so far quite limited.

Another issue is the character of the singularity. Since $a(t)$ tends 
to zero at $t=\pm\infty$ for models with $w\in(-5/3,-1)$, this might suggest 
a sort of Big Bang singularity, though endowed with exotic features. 
However, the sign of the Ricci curvature measured by causal 
geodesics (\ref{ricci}, \ref{ricci1}) prevents this interpretation, 
since in  (\ref{sign}) we see that it is negative (non-focusing) for 
negative $\eta$ in flat models.

In fact, for radial geodesics,
\[\frac{dR}{dt}=\pm\frac{\dot R}{\dot t}=\pm
\frac{P}{a(t)\sqrt{P^2+a^2(t)}}\simeq \pm\frac{1}{a(t)},\]\[
R\simeq R_{0}\pm\frac{1}{c}\frac{t^{1-\eta}}{1-\eta},\] the radial 
coordinate diverges for large $|t|$ in models with $\eta<1$, and shows 
that geodesics are indeed not focusing and therefore geodesics 
diverge instead of converge.

With all these results in mind, we may refine the usual 
classification of singular events in models according to the value of $w$ by introducing 
this new information, bearing in mind that none of these models is 
valid for the whole life of the universe, just for a fraction of it:

\begin{enumerate}
    \item  Events with $\eta>1$, $-1<w<-1/3$: Quintessence models with a Big Bang singularity at $t=0$.

    \item Events with $\eta=1$, $w=-1/3$, $k=-1$: Milne-like models which may 
    have weak or strong singularities at $t=0$  \cite{visser,puiseux}.
    
    \item  Events with $0<\eta<1$, $w>-1/3$: Classical models (dust, 
    radiation\ldots) with a Big Bang singularity at $t=0$.
    
    \item  Events with $\eta=0$: The menagerie of models which are 
    either regular (de Sitter, anti-de Sitter\ldots) or possess 
    sudden, freeze, pressure, higher derivative singularities as 
    described in \cite{visser,classodi,puiseux}, which may be weak 
    or strong.
    
    \item  Events with $-1\le \eta<0$, $w\le-5/3$: Phantom models 
    ranging from $t=-\infty$ to $t=0$ where they meet a Big Rip 
    singularity. Scalar perturbations of these models have been shown 
    to lead to high inhomogeneity, which may prevent the formation of 
    the singularity \cite{fabris}.

    \item  Events with $\eta<-1$, $w\in(-5/3,-1)$: Phantom models with a 
    p.p. curvature singularity at $t=-\infty$ which affects just radial 
    geodesics and a Big Rip singularity at $t=0$ which does not show 
    up for lightlike geodesics.

\end{enumerate}

\section*{Acknowledgments}L.F.-J. is supported by the Spanish Ministry
of Education and Science research grant FIS-2005-05198. The author wishes to 
thank R. Lazkoz, J.M.M. Senovilla and R. Vera for valuable discussions and the 
University of the Basque Country for their hospitality.


\begin{thebibliography}{99} 
\bibitem{supernova}A.G. Riess et al., \textit{Astron.  J.} \textbf{116}, 1009 
(1998); 
S.J. Perlmutter et al., \textit{Astroph.  J.} \textbf{517}, 
565 (1999); 
S.J. Perlmutter et al, \textit{Nature} \textbf{391}, 51 (1998), 
\textit{Bull.  Am.  Astron.  Soc.} \textbf{29}, 1351 (1997); 
P. Garnavich et al, \textit{Astrophys.  J.} \textbf{493}, L53 (1998);
B.P.Schmidt et al, \textit{Astrophys.  J.} \textbf{507}, 46 (1998); 
J.L. Tonry et al., \textit{Astroph.  J.} \textbf{594}, 1 (2003); 
B. Barris et al., \textit{Astroph.  J.} \textbf{602}, 571 (2004); 
R. Knop et al., \textit{Astroph.  J.} \textbf{598} 102 (2003); 
A.G. Riess et al., \textit{Astrophys. J.} \textbf{607}, 665 (2004); 
\bibitem{redshift}ÊN.A. Bahcall et al, \textit{Science} \textbf{284}, 1481 (1999); 
W.J. Percival et al., \textit{Mon. Not.  Roy.  Ast.  Soc.}  \textbf{327}, 1297 (2001); 
M. Tegmark et al., \textit{Phys.  Rev.  D} \textbf{69}, 103501 (2004)
\bibitem{cmbr} D.N. Spergel et al, \textit{Astrophys. J. Suppl.} 
\textbf{148}, 175 (2003) [arXiv:astro-ph/0302209]; 
D. Miller et al., \textit{Astrophys.  J.} \textbf{524}, L1 (1999); 
C. Bennett et al, \textit{Astrophys. J. Suppl.} \textbf{148}, 1 
(2003) [arXiv:astro-ph/0302207]; 
P. de Bernardis et al., \textit{Nature}, \textbf{404}, 955 (2000)  
[arXiv:astro-ph/0004404]; 
S. Hanany et al., \textit{Astrophys.  J.} \textbf{545}, L5 (2000); 
T.J. Pearson et al., \textit{Astrophys. J.} \textbf{591}, 556 (2003); 
B.S. Mason et al., \textit{Astrophys. J.} \textbf{591}, 540 (2003); 
A. Benoit et al., \textit{Astron.  Astrophys.} \textbf{399}, L25 
(2003).

\bibitem{de}
T. Padmanabhan,
\textit{Curr.\ Sci.}\  {\bf 88}, 1057 (2005)  [arXiv:astro-ph/0411044].
\bibitem{caldwell}
R.R. Caldwell,
\textit{Phys.\ Lett.\ B} {\bf 545}, 23 (2002)
[arXiv:astro-ph/9908168].
\bibitem{kahya} E.O. Kahya, V.K. Onemli, arXiv:gr-qc/0612026.
\bibitem{sudden} J.D. Barrow, \textit{Class. Quant. Grav.} {\bf 21}, L79 (2004)
 [arXiv:gr-qc/0403084];  S. Nojiri, S.D. Odintsov, \textit{Phys.\ Lett. B} 
 {\bf 595}, 1 (2004) [arxiv:hep-th/0405078]; 
J.D. Barrow,
  \textit{Class.\ Quant.\ Grav.}\  {\bf 21}, 5619 (2004) 
  [arXiv:gr-qc/0409062]; 
K. Lake,
  \textit{Class.\ Quant.\ Grav.}  {\bf 21}, L129 (2004)
  [arXiv:gr-qc/0407107];  S. Nojiri, S.D. Odintsov, \textit{Phys.\ 
  Rev.\ D} {\bf 70}, 103522 (2004) [arxiv:hep-th/0408170]; 
  M.P. D\c abrowski,
  \textit{Phys.\ Rev.\ D} {\bf 71}, 103505 (2005) 
  [arXiv:gr-qc/0410033]; 
J.D. Barrow, C.G. Tsagas, 
\textit{Class.\ Quant.\ Grav.}\  {\bf 22}, 1563 (2005) 
[arXiv:gr-qc/0411045]; 
L.P. Chimento,  R. Lazkoz,
\textit{Mod.\ Phys.\ Lett.\ A} {\bf 19}, 2479 (2004) 
[arXiv:gr-qc/0405020]; 
 M.P. D\c abrowski,
%
\textit{Phys.\ Lett.\ B} {\bf 625}, 184 (2005)  
[arXiv:gr-qc/0505069]; 
 A. Balcerzak, M.P. D\c abrowski,
\textit{Phys.\ Rev.\ D} {\bf 73}, 101301 (2006)
[arXiv:hep-th/0604034].
\bibitem{mcinnes} B. McInnes, \textit{Class. Quant. Grav.} 
\textbf{24}, 1605 (2007).
\bibitem{visser} C. Catto\"en,  M. Visser,
\textit{Class.\ Quant.\ Grav.}\  {\bf 22}, 4913 (2005) 
[arXiv:gr-qc/0508045]; 
 C. Catto\"en,
MSc thesis, Victoria University of Wellington (2006) 
[arXiv:gr-qc/0606011].
\bibitem{classodi}
S. Nojiri, S.D. Odintsov, S. Tsujikawa,
\textit{Phys.\ Rev.\ D} {\bf 71},  063004 (2005)
[arXiv:hep-th/0501025].
\bibitem{puiseux} L. Fern\'andez-Jambrina, R. Lazkoz, 
\textit{Phys. Rev. D} \textbf{74}, 064030 (2006) [arXiv:gr-qc/0607073].

\bibitem{flrw} L. Fern\'andez-Jambrina,  R. Lazkoz, \textit{Phys. Rev. D}
\textbf{70}, 121503(R) (2004) [arXiv:gr-qc/0410124]
\bibitem{ellis} G.F.R. Ellis, B.G. Schmidt, {Gen. Rel. Grav.}
\textbf{8}, 915 (1977).
\bibitem{tipler} F.J. Tipler, \textit{Phys. Lett.} \textbf{A64}, 8 (1977). 
\bibitem{krolak} A. Kr\'olak, \textit{Class. Quant. Grav.} 
\textbf{3}, 267 (1986). 
\bibitem{dabrowski} M.P. D\c abrowski,  T. Stachowiak, 
M. Szydlowski, 
  \textit{Phys.\ Rev.\ D} {\bf 68}, 103519  (2003). 
  \bibitem{obs}U. Seljak, A. Slosar, P. McDonald, \textit{JCAP} 
  \textbf{0610},  014 (2006) 
    [arXiv:astro-ph/0604335].
\bibitem{HE} S.W. Hawking, G.F.R. Ellis,
\textit{The Large Scale Structure of Space-time}, Cambridge University
Press, Cambridge, (1973).
    
\bibitem{seno} J. M. M. 
Senovilla, \textit{Gen. Rel. Grav.} \textbf{30}, 701, (1998).
\bibitem{eddington} A.S. Eddington, \textit{Nature} \textbf{113}, 192 
(1924);  D. Finkelstein, \textit{Phys. Rev} 
\textbf{110}, 965 (1958).
\bibitem{kruskal} M.D. Kruskal, \textit{Phys. Rev.} \textbf{119}, 
1743 (1960).
\bibitem{chiba} T. Chiba, R. Takahashi, N. Sugiyama, \textit{Class. 
Quant. Grav.} \textbf{22}, 3745 (2005).
\bibitem{hendry} M.P. D\c abrowski, T. Denkiewicz, M.A. Hendry, 
arXiv:astro-ph/07041383.
\bibitem{krorud} A. Kr\'olak, W. Rudnicki, \textit{Int. Journ. Theor. 
Phys.} \textbf{32}, 137 (1993).
\bibitem{rudnicki} W. Rudnicki, R.J. Budzynski, W. Kondracki,
\textit{Mod.\ Phys.\ Lett.\ A} {\bf 21}, 1501 (2006)
[arXiv:gr-qc/0606007].
\bibitem{clarke} C.J.S. Clarke, A. Kr\'olak, \textit{Journ. Geom. Phys.}
\textbf{2}, 17 (1985). 
\bibitem{fabris} J.C. Fabris, S.V.B. Goncalves, \textit{Phys. Rev. D} 
\textbf{74}, 027301 (2006) [arxiv:astro-ph:0603171].
\end{thebibliography}
\end{document}